\documentclass[%
 reprint,
 amsmath,amssymb,
 aps,
 prl
]{revtex4-2}
\usepackage{graphicx}
\usepackage{dcolumn}
\usepackage{bm}
\usepackage{color}
\usepackage[dvipdfmx]{hyperref}

\begin{document}

\preprint{APS/123-QED}

\title{Strong anomalous proximity effect from spin-singlet superconductors}

\author{Satoshi Ikegaya$^{1,2}$, Jaechul Lee$^{3}$, Andreas P. Schnyder$^{1}$, and Yasuhiro Asano$^{3}$}
\affiliation{$^{1}$Max-Planck-Institut f\"ur Festk\"orperforschung, Heisenbergstrasse 1, D-70569 Stuttgart, Germany\\
$^{2}$Department of Applied Physics, Nagoya University, Nagoya 464-8603, Japan\\
$^{3}$Department of Applied Physics, Hokkaido University, Sapporo 060-8628, Japan}
\date{\today}
\begin{abstract}
The proximity effect from a spin-triplet $p_x$-wave superconductor to a dirty normal-metal has been shown to result in various unusual electromagnetic properties,
reflecting a cooperative relation between topologically protected zero-energy quasiparticles and odd-frequency Cooper pairs.
However, because of a lack of candidate materials for spin-triplet $p_x$-wave superconductors, observing this effect has been difficult.
In this paper, we demonstrate that the anomalous proximity effect, which is essentially equivalent to that of a spin-triplet $p_x$-wave superconductor,
can occur in a semiconductor/high-$T_c$ cuprate superconductor hybrid device in which two potentials coexist:
a spin-singlet $d$-wave pair potential and a spin--orbit coupling potential sustaining the persistent spin-helix state.
As a result, we propose an alternative and promising route to observe the anomalous proximity effect
related to the profound nature of topologically protected quasiparticles and odd-frequency Cooper pairs.
\end{abstract}

\maketitle

\section{Introduction}
When a superconductor (SC) is attached to a normal-metal, Cooper pairs (CPs) penetrate into the attached normal segment and modify the electromagnetic properties there.
This phenomenon is known as the proximity effect and has been a central research subject in the field of superconductivity.
When we consider a conventional spin-singlet $s$-wave SC, the attached normal-metal exhibits superconducting-like electromagnetic properties.
However, the proximity effect from a spin-triplet $p_x$-wave SC to a dirty normal-metal (DN) has been shown to result in various counter-intuitive transport properties
such as the quantization of zero-bias conductance (ZBC) in DN/SC junctions~\cite{tanaka_04,tanaka_05(1),asano_07,ikegaya_15,ikegaya_16(1)}
and the fractional current-phase relationship in Josephson currents of SC/DN/SC junctions~\cite{asano_06(1),asano_06(2),ikegaya_16(2)}.
Moreover, although the spin-triplet $p_x$-wave SC shows a diamagnetic response, the magnetic response in the attached DN is reversed to paramagnetic~\cite{asano_11}.
Such a drastic proximity effect of a spin-triplet $p_x$-wave SC has been referred to as an anomalous proximity effect (APE)~\cite{note_ape}.

The APE of spin-triplet $p_x$-wave SCs has attracted intensive attention because its mechanism is related to two particles clad in novel concepts:
a topologically-protected zero-energy quasiparticle and an odd-frequency CP.
The zero-energy states (ZESs) originally located at a surface of a spin-triplet $p_x$-wave SC~\cite{buchholtz_81,hu_94,kashiwaya_00,asano_04,sato_11} can penetrate into the attached DN while retaining the high degree of degeneracy at zero-energy~\cite{tanaka_04,tanaka_05(1),tanaka_05(2),tanaka_07,higashitani_09},
where the robustness of the penetrated ZESs is ensured by topological protection~\cite{ikegaya_15,ikegaya_16(1)}.
The unusual transport properties are a direct consequence of the resonant tunneling of quasiparticles via such topologically protected ZESs in the DN.
Moreover, it has been shown that the ZESs penetrated from a spin-triplet $p_x$-wave are accompanied by
odd-frequency CPs~\cite{tanaka_05(2),tanaka_07,higashitani_09,tanaka_12,asano_13,tamura_19},
which are responsible for the paramagnetic response in the DN~\cite{asano_11,yokoyama_11,buzdin_12,suzuki_14,linder_19}.
The APE is a remarkable phenomenon related to the intrinsic natures of topologically protected zero-energy quasiparticles and odd-frequency CPs.
Thus, experimental observations of this effect are an important topic in the physics of superconductivity.

The main difficulty in observing the APE is a serious lack of candidate materials for spin-triplet $p_x$-wave SCs.
Thus far, several theoretical models for effective $p_x$-wave SCs have been proposed,
including models for semiconductor/$s$-wave SC hybrids under magnetic fields~\cite{alicea_10,you_13,ikegaya_18}
and helical $p$-wave SCs (which are also rare) under magnetic fields~\cite{ikegaya_18,law_13,rosenow_14}.
However, experimentally realizing these models is also challenging because strong Zeeman potentials
that exceed the superconducting pair potentials are needed to induce effective $p_x$-wave superconductivity.
To resolve this stalemate, in this Letter, we explore an alternative route to observing the APE.
\begin{figure}[bbbb]
\begin{center}
\includegraphics[width=0.35\textwidth]{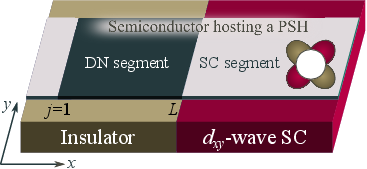}
\caption{Schematic of an effective dirty normal-metal (DN)/$d_{xy}$-wave superconductor (SC) junction in the presence of a persistent spin helix (PSH).}
\label{fig:figure1}
\end{center}
\end{figure}

A central component of our scheme is to demonstrate the APE purely from \textit{spin-singlet} SCs,
whereas the spin-triplet $p_x$-wave SCs have been speculated to be critical for realizing
the APE~\cite{tanaka_04,tanaka_05(1),tanaka_05(2),tanaka_07,higashitani_09,asano_07,ikegaya_15,ikegaya_16(1),asano_06(1),asano_06(2),ikegaya_16(2),asano_11}.
Specifically, we demonstrate the APE in a two-dimensional (2D) semiconductor fabricated on an insulator/high-$T_c$ cuprate SC junction, as shown in Fig.~\ref{fig:figure1}.
We assume a proximity-induced spin-singlet $d_{xy}$-wave pair potential for the segment above the high-$T_c$ cuprate SC.
For the segment above the insulator, we assume a nonmagnetic disorder potential that can be introduced, for example,
using a focused ion beam technique~\cite{muroe_04,stevie_05}.
Consequently, the 2D semiconductor becomes an effective DN/spin-singlet $d_{xy}$-wave SC junction.
In addition, we assume that the semiconductor hosts a persistent spin helix (PSH) state, which has been studied intensively in
the field of spintronics ~\cite{bernevig_06,chang_06,schliemann_17,kohda_17}.
As described in detail later, a spin--orbit coupling (SOC) potential sustaining the PSH induces a spin-triplet $p_x$-wave pairing \textit{correlation}
in the SC segment~\cite{tamura_19(2),asano_20}, whereas the pairing \textit{correlation} does not contribute to the superconducting \textit{gap} directly.
The induced spin-triplet $p_x$-wave pairing correlation in the SC segment is
a source of robust odd-frequency spin-triplet $s$-wave pairing correlation in the attached DN segment.
Moreover, on the basis of an Atiyah--Singer index theorem~\cite{ikegaya_15,ikegaya_16(1),ikegaya_18,ikegaya_17},
we will demonstrate the emergence of topologically-protected ZESs in the DN segment only in the presence of PSH.
Remarkably, PSH states have been already realized in a number of experiments using semiconductor quantum well systems~\cite{awschalom_09,salis_12,kohda_12,salis_14}.
Furthermore, it has been recently shown that the PSH is realized intrinsically in 2D ferroelectric materials such as
group-IV monochalcogenide monolayers~\cite{tsymbal_18,ishii_19,ishii_19(2),picozzi_19,zhao_19,jin_20,nardelli_20,chang_20}.
Thus, the recent rapid progress in the physics of the PSH provides a great potential for the realization of the proposed experimental setup.
Consequently, we report a promising route to observing the APE.

\section{Model}
We describe the present system using a 2D tight-binding model.
A lattice site is indicated by a vector $\boldsymbol{r}=j \boldsymbol{x} + m \boldsymbol{y}$, where $\boldsymbol{x}$ ($\boldsymbol{y}$) is a unit vector in the $x$ ($y$) direction.
The system comprises three segments: a semi-infinite lead wire (ballistic semiconductor segment) for $-\infty \leq j<1$,
a DN segment (dirty semiconductor segment) for $1\leq j \leq L$,
and a semi-infinite SC segment (ballistic semiconductor segment with a proximity-induced pair potential) for $L<j<\infty$.
In the $y$ direction, the number of lattice sites is given by $W$ and a periodic boundary condition is applied.
The Bogoliubov--de Gennes (BdG) Hamiltonian reads $H=H_{\mathrm{kin}}+H_{\mathrm{PSH}}+H_{\Delta}+H_{\mathrm{v}}$.
The kinetic energy is given by
\begin{align}
H_{\mathrm{kin}} =& -t \sum_{\langle \boldsymbol{r}, \boldsymbol{r}^{\prime} \rangle }\sum_{\sigma=\uparrow,\downarrow}
( c_{\boldsymbol{r},\sigma}^{\dagger}c_{\boldsymbol{r}^{\prime},\sigma}+c_{\boldsymbol{r}^{\prime},\sigma}^{\dagger}c_{\boldsymbol{r},\sigma} )
\nonumber\\
& +(4t - \mu) \sum_{\boldsymbol{r}, \sigma} c_{\boldsymbol{r},\sigma}^{\dagger}c_{\boldsymbol{r},\sigma},
\end{align}
where $c_{\boldsymbol{r},\sigma}^{\dagger}$ ($c_{\boldsymbol{r},\sigma}$) is the creation (annihilation) operator for an electron at $\boldsymbol{r}$ with spin $\sigma$;
$t$ and $\mu$ denote the nearest-neighbor hopping integral and chemical potential, respectively.
The PSH is characterized by a unidirectional SOC potential given by
\begin{align}
H_{\mathrm{PSH}} = \frac{i \lambda}{2} \sum_{\boldsymbol{r}, \sigma} s_{\sigma}
(c_{\boldsymbol{r}+\boldsymbol{y},\sigma}^{\dagger}c_{\boldsymbol{r},\sigma}-c_{\boldsymbol{r},\sigma}^{\dagger}c_{\boldsymbol{r}+\boldsymbol{y},\sigma}),
\label{eq:psh}
\end{align}
where $s_{\uparrow (\downarrow)} = +1 (-1)$.
The SOC potential in Eq.~(\ref{eq:psh}) describes a Dresselhaus[110] SOC potential realized in
zinc-blende III--V semiconductor quantum wells grown along the $[110]$ direction~\cite{bernevig_06,chang_06,salis_14}.
The equivalent SOC potentials can also be obtained in quantum wells in which
Rashba and Dresselhaus[100] SOC potentials have equal amplitudes~\cite{bernevig_06,chang_06,awschalom_09,salis_12,kohda_12}
and in ferroelectric thin-film materials~\cite{tsymbal_18,ishii_19,ishii_19(2),picozzi_19,zhao_19,jin_20,nardelli_20,chang_20}.
The proximity-induced spin-singlet $d_{xy}$-wave pair potential is given by
\begin{align}
H_{\Delta}=& \frac{\Delta}{4} \sum_{j=L+1}^{\infty} \sum_{m=1}^{W}
( c_{\boldsymbol{r}+\boldsymbol{x}+\boldsymbol{y},\uparrow}^{\dagger}c_{\boldsymbol{r},\downarrow}^{\dagger}
+c_{\boldsymbol{r},\uparrow}^{\dagger}c_{\boldsymbol{r}+\boldsymbol{x}+\boldsymbol{y},\downarrow}^{\dagger} \nonumber\\
& -c_{\boldsymbol{r}+\boldsymbol{x}-\boldsymbol{y},\uparrow}^{\dagger}c_{\boldsymbol{r},\downarrow}^{\dagger}
-c_{\boldsymbol{r},\uparrow}^{\dagger}c_{\boldsymbol{r}+\boldsymbol{x}-\boldsymbol{y},\downarrow}^{\dagger}) + \mathrm{H.c.},
\end{align}
where $\Delta$ denotes the amplitude of the pair potential.
The disorder potential in the DN segment is described by
\begin{align}
H_{v}=\sum_{j=1}^{L} \sum_{m=1}^{W} \sum_{\sigma} v(\boldsymbol{r}) c_{\boldsymbol{r},\sigma}^{\dagger}c_{\boldsymbol{r},\sigma},
\end{align}
where $v(\boldsymbol{r})$ is given randomly in the range $-X \leq v(\boldsymbol{r}) \leq X$.

In the following numerical calculations, we fix the parameters as $\mu=t$, $\lambda=0.5t$, $\Delta=0.1t$, $L=50$, and $W=50$.
For simplicity, we assume that $t$, $\mu$, and $\lambda$ are uniform in the entire system.
For the ensemble average, 500 samples are used.
To observe the APE experimentally, the thermal coherence length $\xi_T = \sqrt{\hbar D/ 2 \pi k_B T}$ must be longer than the length of the DN segment,
where $T$ and $D$ represent the temperature and the diffusion constant in the DN segment, respectively.
For simplicity, we assume zero temperature in the following calculations.

\section{Results}
\subsection{Anomalous proximity effect}
\begin{figure}[tttt]
\begin{center}
\includegraphics[width=0.34\textwidth]{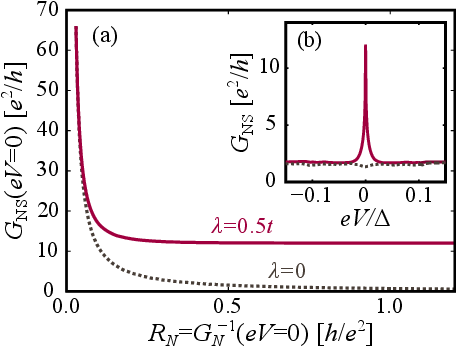}
\caption{(a) ZBC as a function of the normal resistance $R_N$.
(b) Differential conductance at $R_N = 0.55(h/e^2)$ as a function of the bias voltage.
The solid red (dotted black) line denotes the result corresponding to the presence (absence) of the PSH with $\lambda=0.5t$ ($\lambda=0$).}
\label{fig:figure2}
\end{center}
\end{figure}

\begin{figure}[tttt]
\begin{center}
\includegraphics[width=0.5\textwidth]{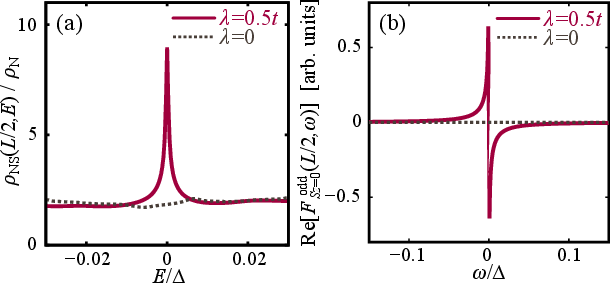}
\caption{(a) LDOS at the center of the DN segment as a function of the energy.
(b) Pair amplitudes for the odd-frequency spin-singlet $s$-wave component at the center of the DN segment as a function of the Matsubara frequency.}
\label{fig:figure3}
\end{center}
\end{figure}

We first focus on the differential conductance in the present device.
Within the Blonder--Tinkham--Klapwijk formalism~\cite{klapwijk_82, bruder_90, kashiwaya_95}, the differential conductance is calculated by
\begin{align}
G_{\mathrm{NS}}(eV) = \frac{e^{2}}{h} \sum_{\zeta,\zeta^{\prime}}
\left[ \delta_{\zeta,\zeta^{\prime}} - \left| r^{ee}_{\zeta,\zeta^{\prime}} \right|^{2}
+ \left| r^{he}_{\zeta,\zeta^{\prime}} \right|^{2} \right]_{E=eV},
\end{align}
where $r^{ee}_{\zeta,\zeta^{\prime}}$ and $r^{he}_{\zeta,\zeta^{\prime}}$ denote a normal and an Andreev reflection coefficient at energy $E$, respectively.
The indexes $\zeta$ and $\zeta^{\prime}$ label an outgoing and incoming channel in the lead wire, respectively.
These reflection coefficients are calculated using recursive Green's function techniques~\cite{fisher_81, ando_91}.
In Fig.~\ref{fig:figure2}(a), we show the ZBC, i.e., $G_{\mathrm{NS}}(eV=0)$ as a function of
the normal resistance $R_N = G^{-1}_{\mathrm{N}}(eV=0)$, where the normal conductance $G_{\mathrm{N}}(eV)$ is calculated by setting $\Delta=0$.
We vary the value of $R_N$ by changing the magnitude of the disorder potential, $X$.
The dotted black line denotes the result corresponding to the absence of the PSH (i.e., $\lambda$=0).
In this case, the ZBC decreases to zero with increasing resistance in the normal segment $R_N$.
Nevertheless, as shown by the solid red line, the ZBC in the presence of the PSH shows saturation with increasing resistance and shows a quantization in the dirty limit as
\begin{align}
\lim_{R_N \rightarrow \infty} G_{\mathrm{NS}}(eV=0) = \frac{2 e^2}{h} |\mathbb{Z}|, \label{eq:quantization}
\end{align}
where $|\mathbb{Z}|=6$ in the case of the present parameters.
In Fig.~\ref{fig:figure2}(b), we also show $G_{\mathrm{NS}}$ at $R_N = 0.55 (h/e^2)$ as a function of the bias voltage.
The results clearly show that only the conductance spectrum in the presence of the PSH exhibits a sharp ZBC peak.
The quantization of the ZBC in the dirty limit is a representative manifestation of the APE~\cite{tanaka_04,tanaka_05(1),ikegaya_15,ikegaya_16(1)}.
The integer number of $\mathbb{Z}$, which characterizes the strength of the APE, will later be derived analytically. 
We now discuss the local density of states (LDOS) in the DN segment.
The LDOS averaged over the lattice sites in the $y$ direction is calculated by
\begin{align}
\rho_{\mathrm{NS}}(j,E) = - \frac{1}{\pi W}
\sum_{m=1}^{W}\mathrm{Tr}\left[ \mathrm{Im} \left\{ \check{G}(\boldsymbol{r},\boldsymbol{r},E+i\delta) \right\}\right],
\end{align}
where $\check{G}(\boldsymbol{r},\boldsymbol{r}^{\prime},E)$ represents the Green's function, $\delta$ is a small imaginary component added to the energy, $E$,
and $\mathrm{Tr}$ denotes the trace in spin and Nambu spaces.
Fig.~\ref{fig:figure3}(a) shows $\rho_{\mathrm{NS}}(j,E)$ at the center of the DN segment (i.e., $j = L/2$) as a function of the energy.
We chose $X=2t$ and $\delta=10^{-4}\Delta$.
The result is normalized by the LDOS in the normal states $\rho_N$ calculated by setting $\Delta=0$.
When the PSH is present, the LDOS exhibits a sharp zero-energy peak (solid red line), whereas the peak is not observed when the PSH is absent (dotted black line).
The zero-energy peak in the LDOS suggests that ZESs originally located at the junction interface penetrate into the DN segment,
which is responsible for the quantization of the conductance minimum in Eq.~(\ref{eq:quantization})~\cite{tanaka_04,tanaka_05(1),ikegaya_15,ikegaya_16(1)}.

Here, we discuss the odd-frequency CPs in the DN segment.
SOC potentials have been shown to induce spin-triplet pairing correlations in spin-singlet SCs~\cite{tamura_19(2),asano_20},
whereas the pairing correlation does not contribute to the superconducting gap directly.
According to the analysis in Ref.~\onlinecite{asano_20}, the SOC potential in Eq.~(\ref{eq:psh}), which sustains the PSH,
generates an even-frequency spin-triplet $p_x$-wave correlation in the spin-singlet $d_{xy}$-wave SC (see also the Supplemental Material~\cite{supplemental}).
Therefore, as in the case of pure DN/$p_x$-wave SC junctions~\cite{tamura_19,asano_20}, we can reasonably expect that
the even-frequency spin-triplet $p_x$-wave pairing correlations induced in the SC segment can function as
the source of odd-frequency spin-triplet $s$-wave correlations in the attached DN segment.
We here focus only on the odd-frequency spin-triplet $s$-wave CPs, whose pair amplitude is evaluated by
\begin{align}
&F^{\mathrm{odd}}_{S_z=0}(j,\omega) = \frac{1}{ \sqrt{2} W}\sum_{m=1}^{W}
[ F^{\mathrm{odd}}_{\uparrow,\downarrow}(\boldsymbol{r},\omega) + F^{\mathrm{odd}}_{\downarrow,\uparrow}(\boldsymbol{r},\omega) ], \\
&F^{\mathrm{odd}}_{\sigma,\sigma^{\prime}}(\boldsymbol{r},\omega) =
[F_{\sigma,\sigma^{\prime}}(\boldsymbol{r},\boldsymbol{r},\omega) - F_{\sigma,\sigma^{\prime}}(\boldsymbol{r},\boldsymbol{r},-\omega)]/2,
\end{align}
where $F_{\sigma,\sigma^{\prime}}(\boldsymbol{r},\boldsymbol{r}^{\prime},\omega)$ represents the anomalous part of the Matsubara Green's function.
We also confirm that other components of the odd-frequency spin-triplet $s$-wave CPs,
\begin{align}
F^{\mathrm{odd}}_{S_z=1(-1)}(j,\omega) = \sum_{m}F^{\mathrm{odd}}_{\uparrow,\uparrow (\downarrow,\downarrow)}(\boldsymbol{r},\omega)/W,
\nonumber
\end{align}
are absent in the present junction.
Fig.~\ref{fig:figure3}(b), shows the real part of $F^{\mathrm{odd}}_{S_z=0}$ at the center of the DN segment as a function of the Matsubara frequency $\omega$,
where the imaginary part of $F^{\mathrm{odd}}_{S_z=0}$ is also found to be zero identically.
As expected, $F^{\mathrm{odd}}_{S_z=0}$ in the DN segment becomes finite in the presence of the PSH (solid red line),
whereas in the absence of the PSH (dotted black line), $F^{\mathrm{odd}}_{S_z=0}=0$.
As a result, we confirm the formation of the odd-frequency spin-triplet $s$-wave CPs in the DN, which is an important aspect of the APE.

\subsection{Index theorem}
We here discuss an Atiyah--Singer index theorem that characterizes the APE in the present junction.
To evaluate the topological property of the SC segment, we remove the DN segment from the system and apply a periodic boundary in the $x$ direction.
Moreover, for simplicity, we describe the present SC in continuous space.
Consequently, the BdG Hamiltonian in momentum space is given by
\begin{gather}
H(\boldsymbol{k}) = \xi(\boldsymbol{k}) \sigma_0 \tau_z 
+ \lambda k_y \hat{\sigma}_z \tau_0 - \Delta(\boldsymbol{k}) \sigma_y \tau_y,
\end{gather}
where $\xi(\boldsymbol{k}) = (\hbar^2 k^2/2 m) - \mu$ with $m$ representing the effective mass of an electron,
$\Delta(\boldsymbol{k}) = \Delta (k_x k_y/k^2_F)$ with $k_F= \sqrt{2m \mu}/\hbar$ representing the Fermi wavenumber,
$\sigma_\alpha$ ($\tau_\alpha$) for $\alpha=x,y,z$ are the Pauli matrices acting on spin (Nambu) space, and $\sigma_0$ ($\tau_0$) is the unit matrix in spin (Nambu) space.
The BdG Hamiltonian $H(\boldsymbol{k})$ intrinsically preserves particle--hole symmetry as $ C H(\boldsymbol{k}) C^{-1}= - H(-\boldsymbol{k})$,
where $C = \tau_x {\cal K}$ with ${\cal K}$ representing the complex conjugation operator.
We also find time-reversal symmetry as $T_- H(\boldsymbol{k}) T_-^{-1}= H(-\boldsymbol{k})$ with $T_- = i\sigma_y \tau_0 {\cal K}$.
Because $C^2=+1$ and $T_-^2=-1$, the BdG Hamiltonian belongs to the DIII symmetry class~\cite{schnyder_08}.
Importantly, because of the nature of the PSH ~\cite{bernevig_06},
the BdG Hamiltonian preserves spin-rotation symmetry along the $z$ axis even in the presence of the SOC potential:
$R_z H(\boldsymbol{k}) R_z^{-1}= H(\boldsymbol{k})$ with $R_z = \sigma_z \tau_z$.
By combining $T_-$ and $R_z$, we obtain $T_+ H(\boldsymbol{k}) T_+^{-1}= H(-\boldsymbol{k})$,
where $T_+ = R_z T_-$ represents an additional time-reversal symmetry obeying $T_+^2 = +1$.
Because $C^2=+1$ and $T_+^2=+1$, we find that $H(\boldsymbol{k})$ can be simultaneously classified into the BDI symmetry class~\cite{schnyder_08}.
The energy spectrum of $H(\boldsymbol{k})$ is given by
$E_{s_\sigma}(\boldsymbol{k}) = \pm \sqrt{\left\{ \xi(\boldsymbol{k}) + s_{\sigma} \lambda k_y \right\}^2 + \Delta^2 (\boldsymbol{k})}$.
The branch of $E_{s_\sigma}(\boldsymbol{k})$ exhibits four superconducting gap nodes at
$(k_x,k_y) = (\pm k_F,  0)$ and $(0, \pm k_{\lambda} - s_{\sigma} \bar{\lambda})$, where
\begin{align}
k_{\lambda} =  \sqrt{k_F^2 + \bar{\lambda}^2}, \quad
\bar{\lambda} = m \lambda / \hbar^2.
\end{align}
On the basis of the Atiyah--Singer index theorem~\cite{ikegaya_15,ikegaya_16(1),ikegaya_18,ikegaya_17},
a topological index that characterizes the number of zero-energy states at a \textit{dirty} surface of a nodal SC is given by
\begin{gather}
\mathbb{Z} = {\sum_{k_y}}^{\prime} w_{\mathrm{BDI}}(k_y), \label{eq:topo_index}\\
w_{\mathrm{BDI}}(k_y) = \frac{i}{4 \pi}
\int dk_x \mathrm{Tr} [ S_{+} \left\{ H (\boldsymbol{k}) \right\}^{-1}
\partial_{k_x} H (\boldsymbol{k}) ], \label{eq:wind}
\end{gather}
where $S_+ = i T_+ C = - \sigma_x \tau_y$ represents the chiral symmetry operator with respect to the BDI symmetry class
and $\sum_{k_y}^{\prime}$ denotes a summation over $k_y$ excluding the nodal points.
Using Eq.~(\ref{eq:wind}), we obtain
\begin{align}
w_{\mathrm{BDI}} (k_y)= \left\{ \begin{array}{cl}
1& \text{for}\quad  k_{\lambda} - \bar{\lambda} <|k_y|< k_{\lambda} + \bar{\lambda}\\
0 & \text{otherwise}
\end{array}\right. ,
\end{align}
and therefore $\mathbb{Z}=\sum_{k_{\lambda} - \bar{\lambda} <|k_y|< k_{\lambda} + \bar{\lambda}}$.
When the momentum $k_y$ is assumed to be a continuous variable, the discrete summation of $k_y$ is replaced with the integration as
\begin{align}
\sum_{k_y} \rightarrow \frac{W}{2 \pi} \int d k_y.
\label{eq:int_ky}
\end{align}
Using Eq.~(\ref{eq:int_ky}), we obtain
\begin{align}
\mathbb{Z} = \left[ (\lambda k_F/2 \mu) N_c \right]_{\mathrm{G}},
\end{align}
where $[ \cdots ]_\mathrm{G}$ is the Gauss symbol giving the integer part of the argument, and $N_c = 2 W k_F/\pi$,
where $[N_c]_\mathrm{G}$ represents the number of propagating channels.
The index $\mathbb{Z}$ becomes finite only in the presence of the PSH ($\lambda \neq 0$).
According to the Atiyah--Singer index theorem~\cite{ikegaya_15,ikegaya_16(1),ikegaya_18,ikegaya_17},
the $|\mathbb{Z}|$ ZESs can robustly remain at zero-energy even in the presence of potential disorders;
they can therefore penetrate into the attached DN while retaining their $|\mathbb{Z}|$-fold degeneracy~\cite{ikegaya_15,ikegaya_16(1)}.
In the presence of the chiral symmetry of $S_+$, each ZES can form a perfect Andreev reflection channel at zero-energy~\cite{ikegaya_16(1), fulga_11},
which explains the ZBC quantization in Eq.~(\ref{eq:quantization}).

We here note that the number of stable ZESs in typical \textit{fully gapped} topological SCs is limited to a few.
However, the present hybrid system can host the ZESs with a high degree of degeneracy at zero-energy.
For instance, when we assume that $\mu = 2\mathrm{meV}$, $\lambda = 10\mathrm{meV}\mathrm{nm}$, and $ m = 0.07 m_e$,
where $m_e$ is the electron rest mass, we obtain $\mathbb{Z} \approx [0.128 N_c]_\mathrm{G}$,
which means that approximately $10\%$ of the propagating channels contribute to the resonant transmission.
We can therefore reasonably expect a drastic signature of the APE, which should be easily detectable in experiments.
\begin{figure}[tttt]
\begin{center}
\includegraphics[width=0.25\textwidth]{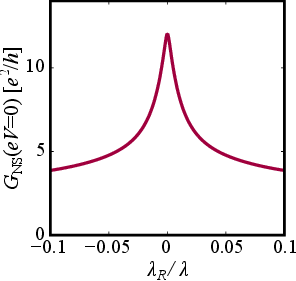}
\caption{ZBC as a function of the strength of the Rashba SOC potential.
We chose $X=2t$.}
\label{fig:figure4}
\end{center}
\end{figure}

\section{Discussion}
We briefly discuss an effect of a perturbative Rashba SOC potential described by $H_R(\boldsymbol{k}) = \lambda_R( k_y \sigma_x \tau_0 - k_x \sigma_y \tau_z)$.
Because the Rashba SOC breaks the chiral symmetry of $S_{+}$, the index $\mathbb{Z}$ in Eq.~(\ref{eq:topo_index}) can no longer be defined.
Therefore, in the presence of Rashba SOC, the ZESs cannot retain their high degree of degeneracy at zero-energy.
Fig.~\ref{fig:figure4} shows the ZBC as a function of the strength of the Rashba SOC $\lambda_R$.
The ZBC is substantially increased in the vicinity of the PSH state (i.e., $\lambda_R=0$).
In principle, the amplitude of Rashba SOC potentials can be tuned experimentally by applying gate voltages or pressures.
Therefore, in experiments, a sudden enhancement in the ZBC as the strength of the Rashba SOC is varied is a possible observable signature of the APE.

In summary, we demonstrate that a spin-singlet $d_{xy}$-wave SC in the presence of PSH exhibits the APE.
The proposed experimental setup can be fabricated by interfacing existing materials.
Our proposal therefore represents a promising approach to observing the APE.

\begin{acknowledgments}
S.I. is supported by a Grant-in-Aid for JSPS Fellows (JSPS KAKENHI Grant No. JP21J00041).
This work is also supported by JSPS KAKENHI (No. JP20H01857), the JSPS Core-to-Core Program (No. JPJSCCA20170002),
the JSPS and Russian Foundation for Basic Research under Japan--Russia Research Cooperative Program Grant No. 19-52-50026,
and the National Science Foundation under Grant No. NSF PHY-1748958.
\end{acknowledgments}

\clearpage
\onecolumngrid
\begin{center}
 \textbf{\large Supplemental Material for \\ ``Strong anomalous proximity effect from spin-singlet superconductors''}\\ \vspace{0.3cm}
Satoshi Ikegaya$^{1,2}$, Jaechul Lee$^{3}$, Andreas P. Schnyder$^{1}$, and Yasuhiro Asano$^{3}$\\ \vspace{0.1cm}
{\itshape $^{1}$Max-Planck-Institut f\"ur Festk\"orperforschung, Heisenbergstrasse 1, D-70569 Stuttgart, Germany\\
$^{2}$Department of Applied Physics, Nagoya University, Nagoya 464-8603, Japan\\
$^{3}$Department of Applied Physics, Hokkaido University, Sapporo 060-8628, Japan}
\date{\today}
\end{center}

In this Supplemental Material, we analyze pairing correlations in the $d_{xy}$-wave superconductor (SC) in the presence of the persistent spin helix (PSH).
The Bogoliubov-de Gennes (BdG) Hamiltonian in momentum space reads
\begin{gather}
\check{H}(\boldsymbol{k})=\left[ \begin{array}{cc}
\hat{h}(\boldsymbol{k}) & \hat{\Delta}(\boldsymbol{k}) \\
-\hat{\Delta}(-\boldsymbol{k}) & -\hat{h}(-\boldsymbol{k})
\end{array} \right], \\
\hat{h}(\boldsymbol{k}) = \xi(\boldsymbol{k}) \hat{\sigma}_0 + \lambda k_y \hat{\sigma}_z,\quad
\hat{\Delta}(\boldsymbol{k}) = \frac{\Delta k_x k_y}{k_F^2} \hat{\sigma}_y,
\end{gather}
which is equivalent to the BdG Hamiltonian in Eq.~(10) of the main text.
This Hamiltonian can be divided into two $2 \times 2$ block components as
\begin{align}
\hat{H}_s (\boldsymbol{k}) = \left[ \begin{array}{cc}
\xi(\boldsymbol{k}) + s \lambda k_y & s\Delta (k_x k_y/k^2_F) \\
s\Delta (k_x k_y/k^2_F) & -\xi(\boldsymbol{k}) - s \lambda k_y
\end{array} \right],
\end{align}
where $s = \pm$.
The Matsubara Green's function in momentum space is obtained by
\begin{align}
\hat{{\cal G}}_s(\boldsymbol{k},\omega) &= \left[ i \omega - \hat{H}_s(\boldsymbol{k}) \right]^{-1} 
= \left[ \begin{array}{cc}
G_s(\boldsymbol{k},\omega) & F_s(\boldsymbol{k},\omega) \\
\underline{F}_s(\boldsymbol{k},\omega) & \underline{G}_s(\boldsymbol{k},\omega) \\
\end{array} \right],
\end{align}
where
\begin{gather}
G_s(\boldsymbol{k},\omega) = -\frac{i\omega+\xi(\boldsymbol{k}) + s \lambda k_y} {A_s (\boldsymbol{k}) }, \quad
\underline{G}_s(\boldsymbol{k},\omega) = -\frac{i\omega - \xi(\boldsymbol{k}) - s \lambda k_y} {A_s (\boldsymbol{k}) }, \\
F_s(\boldsymbol{k},\omega) = \underline{F}_s(\boldsymbol{k},\omega) = \frac{s\Delta (k_x k_y/k^2_F)} {A_s (\boldsymbol{k}) },\\
A_s (\boldsymbol{k}) = \omega^2 + \left\{\xi(\boldsymbol{k}) + s \lambda k_y\right\}^2 + \left\{\Delta (k_x k_y/k^2_F) \right\}^2.
\end{gather}
The anomalous part of the Green's function $F_s(\boldsymbol{k},\omega)$ is deformed as
\begin{gather}
F_s(\boldsymbol{k},\omega) = s F_{d_{xy}}(\boldsymbol{k},\omega) + F_{p_x}(\boldsymbol{k},\omega), \label{eq:supp_pair}\\
F_{d_{xy}}(\boldsymbol{k},\omega) =
\frac{\Delta B(\boldsymbol{k})}{B(\boldsymbol{k})^2 - 4 \lambda^2 \xi^2(\boldsymbol{k}) k^2_y} \frac{k_x k_y}{k^2_F}, \quad
F_{p_x}(\boldsymbol{k},\omega) =
- \frac{2 \Delta \lambda \xi(\boldsymbol{k}) k_y }{B(\boldsymbol{k})^2 - 4 \lambda^2 \xi^2(\boldsymbol{k}) k^2_y} \frac{k_x k_y}{k^2_F},\\
B(\boldsymbol{k}) = \omega^2 + \xi^2(\boldsymbol{k}) + \lambda^2 k^2_y + \left\{\Delta (k_x k_y/k^2_F) \right\}^2.
\end{gather}
The first term in Eq.~(\ref{eq:supp_pair}) describes the $d_{xy}$-wave pairing correlation satisfying
\begin{align}
F_{d_{xy}}(-k_x, k_y, \omega) = F_{d_{xy}}(k_x, -k_y, \omega) = - F_{d_{xy}}(k_x, k_y, \omega), \quad
F_{d_{xy}}(-k_x, -k_y, \omega) = F_{d_{xy}}(k_x, k_y, \omega).
\end{align}
On the other hand, the second term in Eq.~(\ref{eq:supp_pair}) describes the induced $p_{x}$-wave pairing correlation satisfying
\begin{align}
F_{p_{x}}(-k_x, k_y, \omega) = - F_{p_{x}}(k_x, k_y, \omega), \quad
F_{p_{x}}(k_x, -k_y, \omega) = F_{p_{x}}(k_x, k_y, \omega),
\end{align}
where we note that $F_{p_x}(\boldsymbol{k},\omega)=0$ in the absence of the PSH (i.e., $\lambda = 0$).
In the original basis, the anomalous part of the Green's function is given by
\begin{align}
\hat{F}(\boldsymbol{k},\omega)  = \left[ \begin{array}{cc}
F_{\uparrow, \uparrow}(\boldsymbol{k},\omega) & F_{\uparrow, \downarrow}(\boldsymbol{k},\omega) \\
F_{\downarrow, \uparrow}(\boldsymbol{k},\omega) & F_{\downarrow, \downarrow}(\boldsymbol{k},\omega) \\
\end{array} \right] 
=\left[ \begin{array}{cc}
0 &  F_{d_{xy}}(\boldsymbol{k},\omega) + F_{p_x}(\boldsymbol{k},\omega)\\
-F_{d_{xy}}(\boldsymbol{k},\omega) + F_{p_x}(\boldsymbol{k},\omega) & 0 \\
\end{array} \right].
\end{align}
As also discussed in the main text, the induced spin-triplet $p_{x}$-wave pairing correlation $F_{p_x}(\boldsymbol{k},\omega)$ can be the source of
the odd-frequency spin-triplet $s$-wave pairing correlation in a dirty normal metal attached to the SC.
\end{document}